\documentclass[%
reprint,
superscriptaddress,
amsmath,amssymb,aps,prl,
]{revtex4-2}

\usepackage{graphicx}
\usepackage{dcolumn}
\usepackage{bm}
\usepackage{xcolor}%

\usepackage[colorlinks = true, linkcolor = blue, citecolor = blue, urlcolor = blue]{hyperref}

\renewcommand{\selectlanguage}[1]{}

\newcommand{\n}[1]{\mathrm{#1}} 
\newcommand{\im}[1]{\mathit{#1}} 
\newcommand{\mc}[1]{\mathcal{#1}} 

\newcommand{\refb}[1]{\textcolor{black}{#1}}
\newcommand{\refa}[1]{\textcolor{black}{#1}}
\newcommand{\refc}[1]{\textcolor{black}{#1}}

\def\Re{$^{187}$Re}
\def\Ho{$^{163}$Ho}
\def\Er{$^{162}$Er}
\def\Hom{$^{166m}$Ho}

\def\mum{$\mu$m}

\def\mus{$\mu$s}

\def\Ho{$^{163}$Ho}

\def\Er{$^{162}$Er}

\def\psi{Paul Scherrer Institut (PSI), Villigen, Switzerland}

\begin{document}


\title{Most stringent bound on electron neutrino mass obtained with a scalable low temperature microcalorimeter array}

\author{B.K.~Alpert}
\affiliation{National Institute of Standards and Technology (NIST), Boulder, Colorado, USA}

\author{M.~Balata}
\affiliation{Laboratori Nazionali del Gran Sasso (LNGS), INFN, Assergi (AQ), Italy}

\author{D.T.~Becker}
\affiliation{University of Colorado, Boulder, Colorado, USA}

\author{D.A.~Bennett}
\affiliation{National Institute of Standards and Technology (NIST), Boulder, Colorado, USA}

\author{M.~Borghesi}
\email{matteo.borghesi@unimib.it}
\affiliation{Dipartimento di Fisica, Universit\`a di Milano-Bicocca, Milano, Italy}
\affiliation{Istituto Nazionale di Fisica Nucleare (INFN), Sezione di Milano-Bicocca, Milano, Italy}

\author{P.~Campana}
\affiliation{Dipartimento di Fisica, Universit\`a di Milano-Bicocca, Milano, Italy}
\affiliation{Istituto Nazionale di Fisica Nucleare (INFN), Sezione di Milano-Bicocca, Milano, Italy}

\author{R.~Carobene}
\affiliation{Dipartimento di Fisica, Universit\`a di Milano-Bicocca, Milano, Italy}
\affiliation{Istituto Nazionale di Fisica Nucleare (INFN), Sezione di Milano-Bicocca, Milano, Italy}

\author{M.~De~Gerone}
\affiliation{Istituto Nazionale di Fisica Nucleare (INFN), Sezione di Genova, Genova, Italy}

\author{W.B.~Doriese}
\affiliation{National Institute of Standards and Technology (NIST), Boulder, Colorado, USA}

\author{M.~Faverzani}
\affiliation{Dipartimento di Fisica, Universit\`a di Milano-Bicocca, Milano, Italy}
\affiliation{Istituto Nazionale di Fisica Nucleare (INFN), Sezione di Milano-Bicocca, Milano, Italy}

\author{L.~Ferrari~Barusso}
\affiliation{Dipartimento di Fisica, Universit\`a di Genova, Genova, Italy}
\affiliation{Istituto Nazionale di Fisica Nucleare (INFN), Sezione di Genova, Genova, Italy}

\author{E.~Ferri}
\affiliation{Istituto Nazionale di Fisica Nucleare (INFN), Sezione di Milano-Bicocca, Milano, Italy}

\author{J.W.~Fowler}
\affiliation{National Institute of Standards and Technology (NIST), Boulder, Colorado, USA}

\author{G.~Gallucci}
\affiliation{Istituto Nazionale di Fisica Nucleare (INFN), Sezione di Genova, Genova, Italy}

\author{S.~Gamba}
\affiliation{Dipartimento di Fisica, Universit\`a di Milano-Bicocca, Milano, Italy}
\affiliation{Istituto Nazionale di Fisica Nucleare (INFN), Sezione di Milano-Bicocca, Milano, Italy}

\author{J.D~Gard}
\affiliation{University of Colorado, Boulder, Colorado, USA}

\author{F.~Gatti}
\affiliation{Dipartimento di Fisica, Universit\`a di Genova, Genova, Italy}
\affiliation{Istituto Nazionale di Fisica Nucleare (INFN), Sezione di Genova, Genova, Italy}

\author{A.~Giachero}
\affiliation{Dipartimento di Fisica, Universit\`a di Milano-Bicocca, Milano, Italy}
\affiliation{Istituto Nazionale di Fisica Nucleare (INFN), Sezione di Milano-Bicocca, Milano, Italy}

\author{M.~Gobbo}
\affiliation{Dipartimento di Fisica, Universit\`a di Milano-Bicocca, Milano, Italy}
\affiliation{Istituto Nazionale di Fisica Nucleare (INFN), Sezione di Milano-Bicocca, Milano, Italy}

\author{U.~Köster}
\affiliation{Institut Laue-Langevin (ILL), Grenoble, France}

\author{D.~Labranca}
\affiliation{Dipartimento di Fisica, Universit\`a di Milano-Bicocca, Milano, Italy}
\affiliation{Istituto Nazionale di Fisica Nucleare (INFN), Sezione di Milano-Bicocca, Milano, Italy}

\author{M.~Lusignoli}
\affiliation{Istituto Nazionale di Fisica Nucleare (INFN), Sezione di Roma 1, Roma, Italy}
\affiliation{Dipartimento di Fisica, Sapienza, Universit\`a di Roma, Roma, Italy}

\author{P.~Manfrinetti}
\affiliation{Dipartimento di Chimica, Universit\`a di Genova, Genova, Italy}

\author{J.A.B.~Mates}
\affiliation{National Institute of Standards and Technology (NIST), Boulder, Colorado, USA}

\author{E.~Maugeri}
\affiliation{Paul Scherrer Institut (PSI), Villigen, Switzerland}

\author{R.~Moretti}
\affiliation{Dipartimento di Fisica, Universit\`a di Milano-Bicocca, Milano, Italy}
\affiliation{Istituto Nazionale di Fisica Nucleare (INFN), Sezione di Milano-Bicocca, Milano, Italy}

\author{S.~Nisi}
\affiliation{Laboratori Nazionali del Gran Sasso (LNGS), INFN, Assergi (AQ), Italy}

\author{A.~Nucciotti}
\email{angelo.nucciotti@unimib.it}
\affiliation{Dipartimento di Fisica, Universit\`a di Milano-Bicocca, Milano, Italy}
\affiliation{Istituto Nazionale di Fisica Nucleare (INFN), Sezione di Milano-Bicocca, Milano, Italy}

\author{G.C.~O’Neil}
\affiliation{National Institute of Standards and Technology (NIST), Boulder, Colorado, USA}

\author{L.~Origo}
\affiliation{Dipartimento di Fisica, Universit\`a di Milano-Bicocca, Milano, Italy}
\affiliation{Istituto Nazionale di Fisica Nucleare (INFN), Sezione di Milano-Bicocca, Milano, Italy}

\author{G.~Pessina}
\affiliation{Istituto Nazionale di Fisica Nucleare (INFN), Sezione di Milano-Bicocca, Milano, Italy}

\author{S.~Ragazzi}
\affiliation{Dipartimento di Fisica, Universit\`a di Milano-Bicocca, Milano, Italy}
\affiliation{Istituto Nazionale di Fisica Nucleare (INFN), Sezione di Milano-Bicocca, Milano, Italy}

\author{C.D.~Reintsema}
\affiliation{National Institute of Standards and Technology (NIST), Boulder, Colorado, USA}

\author{D.R.~Schmidt}
\email{dan.schmidt@nist.gov}
\affiliation{National Institute of Standards and Technology (NIST), Boulder, Colorado, USA}

\author{D.~Schumann}
\affiliation{Paul Scherrer Institut (PSI), Villigen, Switzerland}

\author{D.S~Swetz}
\affiliation{National Institute of Standards and Technology (NIST), Boulder, Colorado, USA}

\author{Z.~Talip}
\affiliation{Paul Scherrer Institut (PSI), Villigen, Switzerland}

\author{J.N.~Ullom}
\email{joel.ullom@nist.gov}
\affiliation{National Institute of Standards and Technology (NIST), Boulder, Colorado, USA}

\author{L.R.~Vale}
\affiliation{National Institute of Standards and Technology (NIST), Boulder, Colorado, USA}

\date{\today}

\begin{abstract}
The determination of the absolute neutrino mass scale remains a fundamental open question in particle physics, with profound implications for both the Standard Model and cosmology. Direct kinematic measurements, independent of model-dependent assumptions, provide the most robust approach to address this challenge.
In this Letter, we present the most stringent upper bound on the effective electron neutrino mass ever obtained with a calorimetric measurement of the electron capture decay of $^{163}$Ho. The HOLMES experiment employs an array of ion-implanted transition-edge sensor (TES) microcalorimeters, achieving an average energy resolution of 6\,eV FWHM with a scalable, multiplexed readout technique. With a total of $7\times10^7$ decay events recorded over two months and a Bayesian statistical analysis, we derive an upper limit of $m_{\beta}<27$\,eV/c$^2$ at 90\% credibility.
These results validate the feasibility of $^{163}$Ho calorimetry for next-generation neutrino mass experiments and demonstrate the potential of a scalable TES-based microcalorimetric technique to push the sensitivity of direct neutrino mass measurements beyond the current state of the art.
\end{abstract}

\maketitle 
Measuring the mass of neutrinos or antineutrinos is one of the last critical tasks that need attention to complete the understanding of the Standard Model of elementary particles and their interactions.
While next-generation neutrino experiments are expected to tackle the mass-ordering problem \cite{abusleme_potential_2025,thedunecollaboration_long-baseline_2020}, and the neutrinoless double beta decay searches probe the Majorana nature of neutrinos \cite{kamland-zencollaboration_search_2023,adams_search_2022b,agostini_final_2020,anton_search_2019}, only direct neutrino mass experiments can provide the definitive answer on the absolute mass scale.
\refb{Additionally, increasing tensions with the results of oscillation experiments make the neutrino mass derived from cosmological observations, analyzed within the framework of the $\Lambda$CDM model and its extensions, less reliable \cite{gariazzo_quantifying_2023,wang_updating_2024}.}
\refb{The strength of direct neutrino mass experiments is that they rely solely on the conservation of energy and momentum in weak nuclear beta decays to determine the neutrino mass observable which, for current instruments, is approximated by the effective (anti)neutrino mass $m_{\beta} = \sqrt{\sum_{i=1,2,3} |U_{ei}^2| m_i^2}$, where $U_{ei}$ are the elements of the first row of the PMNS matrix and $m_i$ are the masses of each neutrino mass eigenstate.}
\refb{Generally, the signature of the neutrino mass is identified by a corresponding reduction in the total kinetic energy available to detectable particles, which in turn modifies the spectral shape in the vicinity of the endpoint.}
To date, experiments studying tritium beta decay have provided the most stringent limits on the antineutrino mass. The latest of these experiments, KATRIN, leverages magnetic adiabatic collimation with an electrostatic filter (MAC-E filter) to analyze the electrons emitted by a gaseous tritium source. KATRIN is currently taking data and is approaching its planned sensitivity of about 300 meV on the antineutrino mass \cite{aker_direct_2022}. At the same time, KATRIN is reaching the limit of its technique, and further improvements in the sensitivity of direct measurements require radically new developments.
One such development is the KATRIN++ project, which aims to enhance KATRIN’s sensitivity by adopting low-temperature detectors for electron energy differential spectroscopy \cite{kovac_comparison_2025}.
Proposed new techniques involve the use of Cyclotron Resonance Electron Spectroscopy (CRES) alone (Project8 \cite{esfahani_determining_2017} and QTNM \cite{amad_determining_2024}) or in combination with more advanced electron filtering techniques (PTOLEMY \cite{betti_neutrino_2019}), however, they are all at an embryonic stage. The most advanced of these, Project8, has recently achieved a sensitivity of about 150\,eV on the antineutrino mass \cite{project8collaboration_tritium_2023}.

An alternative experimental approach is provided by low-temperature microcalorimetry \cite{nucciotti_use_2016b}.
In this method, the decaying radionuclides are embedded within the absorber of low-temperature detectors, which are typically hundreds of microns in size.
This configuration allows for high-resolution spectroscopy of the total energy released during the decay process, except for the portion carried away by neutrinos.
Compared to integral spectrometry with MAC-E filters, this approach eliminates uncertainties related to the decay final states.
Additionally, it detects decays with nearly 100\% efficiency and, by measuring the spectrum in parallel in each detector, optimizes the usage of measuring time. These advantages together enable a faster accumulation of statistics.
Furthermore, due to the fundamentally different systematic uncertainties \cite{nucciotti_statistical_2014b,nucciotti_expectations_2010a}, calorimetry and the use of radioactive isotopes other than tritium make these experiments an ideal complement for strengthening the robustness of direct neutrino mass measurements.

First attempts involved the use of \Re\ as the beta decaying isotope and achieved sensitivities around 20\,eV on the antineutrino mass \cite{arnaboldi_bolometric_2003b,sisti_new_2004,gatti_microcalorimeter_2001}, but the lack of scalability ultimately led to the abandonment of more ambitious experimental plans.
Subsequently, several new projects (HOLMES \cite{alpert_holmes_2015b}, ECHo \cite{gastaldo_electron_2017}, and NUMECS \cite{croce_development_2016}) started to study the electron capture of \Ho\ as proposed in \cite{derujula_calorimetric_1982}.

\refb{In this letter, we present the first physics result of the HOLMES experiment \cite{alpert_holmes_2015b}, which improves upon the result from ECHo in \cite{velte_high-resolution_2019} and establishes the calorimetric technique as a highly mature and promising approach for advancing direct neutrino mass sensitivity.}
Future sensitive \Ho\ based neutrino mass experiments have the additional compelling potential to give valuable insights into differences between the neutrino and antineutrino masses, which would indicate CPT violation and have profound implications for the understanding of fundamental physics, as it would challenge the Standard Model and suggest new physics beyond it \cite{moura_searches_2022}.

\begin{figure}[tb!]
\includegraphics[width=0.5\textwidth]{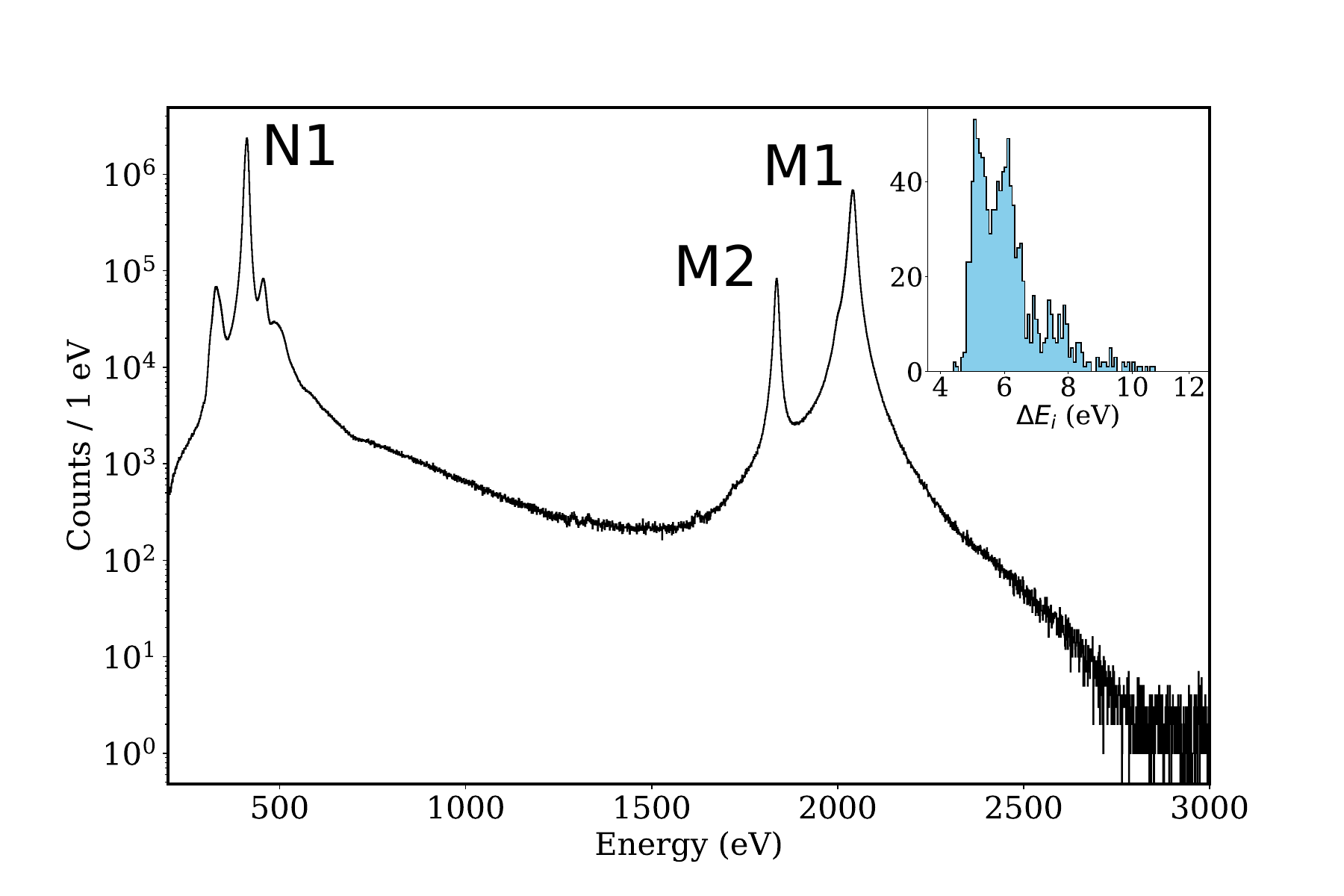}
\caption{\label{fig:spettro}
The total recorded \Ho\ calorimetric spectrum obtained summing about 1000 partial calibrated spectra measured with the HOLMES microcalorimeters.
The spectrum contains about $6\times10^7$\,events above the 300\,eV threshold. The top-right inset shows the distribution of the energy resolution (FWHM) of the individual partial spectra, evaluated from the noise equivalent power (NEP). \refc{The observed double-peaked structure in this distribution reflects an improvement in detector performance between the two physics runs.}}
\end{figure}

A calorimetric neutrino mass experiment using \Ho\ measures the energy released – primarily through electrons \footnote{The total fluorescence yield is expected to be of the order of $10^{-4}$} – following the electron capture (EC) decay:
\begin{equation}
^{163}\mathrm{Ho} + e^- \rightarrow ^{163}\mathrm{Dy} + \nu_e,
\end{equation}
which features the lowest known $Q$-value (about 2863\,eV \cite{schweiger_penning-trap_2024b}) and a half-life of approximately 4750\,years, much shorter than that of \Re, thereby yielding a higher specific activity that is more suitable for use in microcalorimeters.
\refb{The \Ho\ calorimetric spectrum, shown in Fig.\,\ref{fig:spettro}, has as its outstanding feature a combination of Breit-Wigner shaped peaks corresponding to the binding energies of atomic electrons that can be captured (e.g., electrons in the $3s$ shell, M1, or higher shells with binding energies below the $Q$-value, as allowed by energy conservation, and with non-vanishing wave functions at the nucleus).}
Additional, fainter contributions are given by shake-up and shake-off atomic rearrangements following higher order excitations \cite{robertson_examination_2015,derujula_calorimetric_2016,faessler_neutrino_2017,brass_initio_2020}. Although the full spectral shape is non-trivial and a full analytical description is still lacking, the region of interest (ROI) for the neutrino mass estimation – the endpoint of the spectrum – is remarkably smooth and shaped mostly by the phase space singularity. The lack of features at the endpoint is, indeed, a strong advantage of the calorimetric approach.
The endpoint of the \Ho\ spectrum is dominated by the right wing of the M1 peak at about 2041\,eV and the exponential tail of the highest-energy shake-off.

\begin{figure}[tb!]
\includegraphics[width=0.45\textwidth]{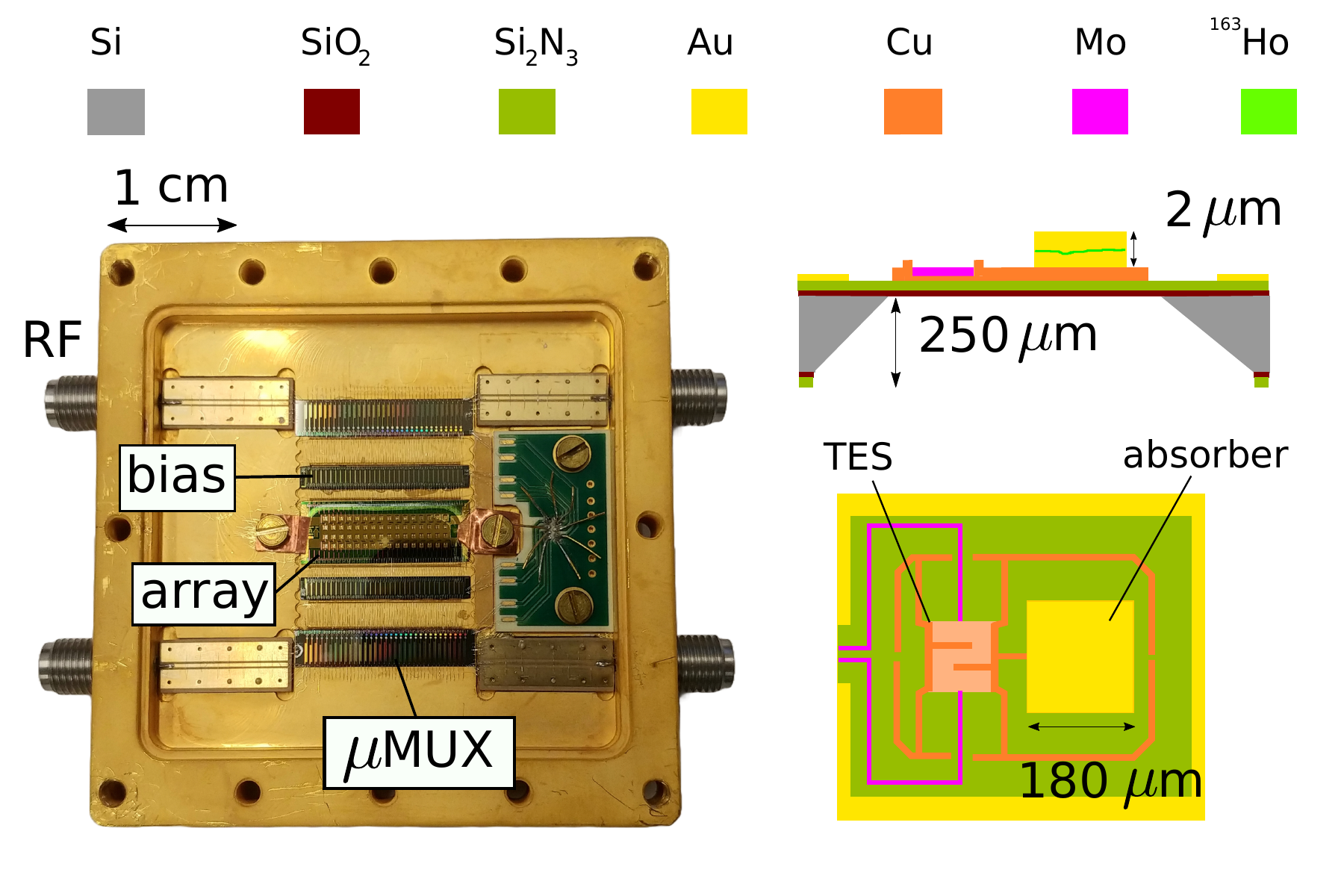}
\caption{\label{fig:holder}
Left: Copper box containing the 64 TES array in the middle. The two chips on either sides of the array are the bias network and the microwave multiplexer, respectively. The array dimensions are approximately ($20\times10$)\,mm$^2$. The multiplexer has the feedline aligned with the SMA connectors used for feeding the readout tones. For readout, two SMAs on one side are connected via a short semirigid coaxial cable. Right: Schematic, not to scale, representation of the HOLMES TES microcalorimeter used in the experiment.}
\end{figure}

HOLMES uses arrays of Transition Edge Sensor (TES) microcalorimeters \cite{alpert_high-resolution_2019b} operated at a temperature of about 95\,mK in a $^3$He/$^4$He dilution refrigerator. The \Ho\  nuclei are ion-implanted at a shallow depth of approximately 100\,\AA{} in a ($180\times180$)\,\mum$^2$ gold layer (see Appendix A for details on isotope preparation and implantation) which is then covered with the deposition of a second overlapping gold layer.
Both layers constitute the absorber of the microcalorimeter and each has a thickness of approximately 1\,\mum, ensuring the full absorption of the radiation emitted in the decay, as required for a calorimetric measurement. The absorber is strongly thermally coupled to the TES sensor, allowing the detection of temperature variations induced by $^{163}$Ho decays.
\refb{
After the \Ho\ source is embedded, the SiN membranes that determine the thermal conductance between the TESs and the heat bath are released \cite{borghesi_updated_2023a}.  The completed array is then mounted in the gold-plated copper box shown in Fig.\,\ref{fig:holder}.  Electrical connections to the devices are made using aluminum wirebonds, and thermal connections between the array and the copper box are made using a combination of gold wirebonds and beryllium-copper clips.
}

The experiment presented here employs an array of 64 TES microcalorimeters. These microcalorimeters are arranged in a $16\times4$ matrix, as shown in Fig.\,\ref{fig:holder}, and their signals are frequency-multiplexed in the (4-8)\,GHz band, leveraging non-hysteretic rf-SQUIDs as current-to-frequency transducers, linearized through flux ramp modulation \cite{becker_working_2019b}.
The multiplexed signals are recovered at room temperature using a heterodyne readout scheme (further technical details on the readout electronics can be found in Appendix B) and software-triggered pulses are stored on disk for further offline processing \cite{borghesi_first_2022}.

This microcalorimeter array, together with the multiplexed readout configuration, forms the foundational building block of our research program. The goal is to further develop and scale the current prototype in the coming years, ultimately leading to an experiment with enhanced statistical sensitivity in the sub-eV range.

When no \Ho\ is implanted, the multiplexed HOLMES microcalorimeters designed, fabricated and measured as described above show an energy resolution of about 4 to 5\,eV FWHM on the 6\,keV X-ray line of manganese. Signals have an approximately double exponential shape with rise and decay times of about 20\,$\mu$s and 600\,\mus, respectively.

In this Letter we report the results of the analysis of the data collected in two physics runs, lasting 2 months for a total live time exposure of about $7\times 10^4$\,detector$\times$hour with about $7\times 10^7$\,\Ho\ decays.
Following the implantation, 48 detectors were found to exhibit non-zero activity, reaching up to 0.6\,Bq, with an average activity of 0.27\,Bq. 
\refb{The remaining detectors showed activity levels too low to allow for reliable calibration (see Appendix\,C for details).}
The total activity of the array is approximately 15\,Bq, corresponding to a source of about $3.2\times10^{12}$\,\Ho\ nuclei, i.e. about 0.9\,ng.

The final spectrum analyzed (Fig.\,\ref{fig:spettro}) for the neutrino mass is obtained by adding the spectra of the 48 active detectors, which, in turn, are obtained by joining their energy-calibrated partial spectra from data collected during periods lasting from 2 to 5\,days.
Additional information about the procedure adopted to calibrate the partial spectra are given in Appendix C.

The energies of the most prominent peaks in the \Ho\ calorimetric spectrum – labeled as M1 (2040.8(3)\,eV), M2 (1836.4(8)\,eV) and N1(411.7(1)\,eV) in Fig.\,\ref{fig:spettro} – were measured during a dedicated run, in which the detectors were exposed to an X-ray source emitting the K lines of aluminum \cite{schweppe_accurate_1994} and chlorine \cite{bearden_x-ray_1967,krause_natural_1979}. The positions of the peaks are in very good agreement with those found in \cite{ranitzsch_characterization_2017}.
Since the amplitude of the signals of a TES is a slightly non-linear estimator of the energy of the event, the energy calibration of the raw \Ho\ spectra is achieved by extrapolating the positions of the M1, M2 and N1 peaks with a quadratic binomial.

The energy resolutions measured in the calibrated spectra correlate as expected with the additional heat capacity introduced by \Ho\ in the absorbers. They are primarily determined by the intrinsic detector noise and progressively degrade with increasing activities. The FWHM resolutions of all partial spectra have an average value of $\langle \Delta E_\n{FWHM}\rangle = (6 \pm 1)$\,eV, with a minimum of approximately 4.4\,eV.

The final spectrum, shown in Fig.\,\ref{fig:spettro} contains about $6\times10^7$\,events above the common analysis threshold set to about 300\,eV and can be described by the expression
\begin{equation}
\label{eq:SpecSumTrue}
\mc{S}_\n{exp} = \sum_i  [N_i (\mc{S}_\n{Ho} +f^{pp}_i \mc{S}_\n{Ho}^{pp}) + \mc{B}_i] * \mc{R}_i
\end{equation}
where $\mc{S}_\n{Ho}$ is the true calorimetric EC energy spectral distribution in the calorimetric energy $E_c$ and the summation is carried over all the calibrated partial spectra.
$\mc{S}_\n{Ho}^{pp}(E_c)$ is the true pile-up spectrum accounting for time unresolved \Ho\ decays: it is given by the self-convolution of the calorimetric EC decay spectrum $\mc{S}_\n{Ho}$ and extends up to twice the endpoint energy \cite{nucciotti_statistical_2014b}. In first approximation, these events have a probability of $f_i^{pp} = \tau^\n{R}_i A_i$, where, for each of the $i-$th calibrated spectra,  $\tau^\n{R}_i$ and $A_i$ are the detector time resolution\,\footnote{We find that the time resolution of our detectors is completely independent of the implanted activity. It is primarily determined by the signal sampling time, approximately a few \mus, which is therefore better than the signal rise time.} and implanted \Ho\ activity, respectively.
For the detectors of this work $f_i^{pp} \lesssim 10^{-5}$, thus making the contribution of the pile-up component in Eq.\,(\ref{eq:SpecSumTrue}) negligible.
$N_i$ are normalization factors taking care of the \Ho\ decays in each spectrum and
$\mc{B}_i(E_c)$ are the energy distributions of spurious events caused by the environmental radioactivity and cosmic rays which are estimated to be flat in the ROI \cite{borghesi_background_2024}.
Finally, in Eq.\,(\ref{eq:SpecSumTrue}) the sum of the true spectra is convolved with detector energy response function $\mc{R}_i(E_c)$ of $i-$th calibrated spectrum, which from detector characterization turns out to be simply Gaussian with FWHM $\Delta E_i$.


\refb{Applying the properties of convolution and for a constant background term, Eq.\,(\ref{eq:SpecSumTrue}) can be rewritten as
\begin{equation}
\label{eq:SpecSum}
\mathcal{S}_{\mathrm{exp}}
= \bigl[N_{\mathrm{tot}}\,( \mathcal{S}_{\mathrm{Ho}} + f^{pp}_{\mathrm{eff}}\,\mathcal{S}_{\mathrm{Ho}}^{pp})\bigr]
\ast \mathcal{R}_{\mathrm{eff}} + b_{\mathrm{eff}}
\end{equation}
with $N_{\mathit{tot}} =  \sum_i N_i$.
Although each true response $\mathcal{R}_i(E_c)$ is Gaussian, their weighted sum is not exactly Gaussian. However, for $\mathcal{O}(10^3)$ spectra with similar FWHM, the composite response converges to a single Gaussian 
$\mathcal{R}_{\rm eff}(E_c)\simeq\mathcal{G}\bigl(E_c\mid0,\Delta E_{\rm eff}\bigr)$
of effective width \(\Delta E_{\rm eff}\), which we leave free in the fit. The response \(\mathcal{R}_{\rm eff}\) deviates in shape from \(\sum_i N_i\,\mathcal{R}_i\) in Eq.\,(\ref{eq:SpecSumTrue}) by less than 2\% within \(\pm\Delta E_{\rm eff}\). Dedicated Monte\,Carlo studies show that, given our current statistics, this substitution does not bias the fit (see Appendix\,E for details).
Since $f^{pp}_{\mathrm{eff}}\,\mathcal{S}_{\mathrm{Ho}}^{pp}$ is subdominant and smooth, the same Gaussian approximation applies without affecting the fit.}
An additional implicit approximation already applied in Eq.\,(\ref{eq:SpecSumTrue}) is the assumption of a perfectly linearized energy response of the detectors. However, the adopted quadratic binomial calibration remains an approximation which, when extrapolated to the ROI beyond the three interpolated calibration points (M1, M2, and N1), may introduce a non-trivial systematic distortion in the summed spectrum of Eq.\,(\ref{eq:SpecSum}).
Applying the same three point calibration procedure to the measurements with the external calibration source, the residual nonlinearity in the detector energy responses is measured to cause deviations $<1\%$ on the chlorine K$\alpha$ positions at about 2600\,eV.
Monte Carlo simulations demonstrate that the impact of all the above approximations on neutrino mass estimation is negligible compared to the current statistical fluctuations \refa{(see Appendix\,E for more details)}.

To perform a sensitive neutrino mass estimation with \Ho, the ROI must be chosen \textit{cum grano salis}.
The upper energy limit should extend beyond the expected EC end point ($E_0$) to constrain the background count rate per detector which is found to be $(1.7\pm0.1)\times10^{-4}$/eV/day between 2900\,eV and 3500\,eV, consistent with the expectations \cite{borghesi_background_2024}.
The choice of the low energy limit, on the other hand, is a trade off. It must be low enough to allow for a precise statistical estimation of both $m_{\beta}$ and $E_0$, yet close enough to the endpoint to ensure that the assumption of spectral smoothness holds, thereby enabling the description of \Ho\ with only a few simple terms.

\begin{figure}[tb!]
\includegraphics[width=0.45\textwidth]{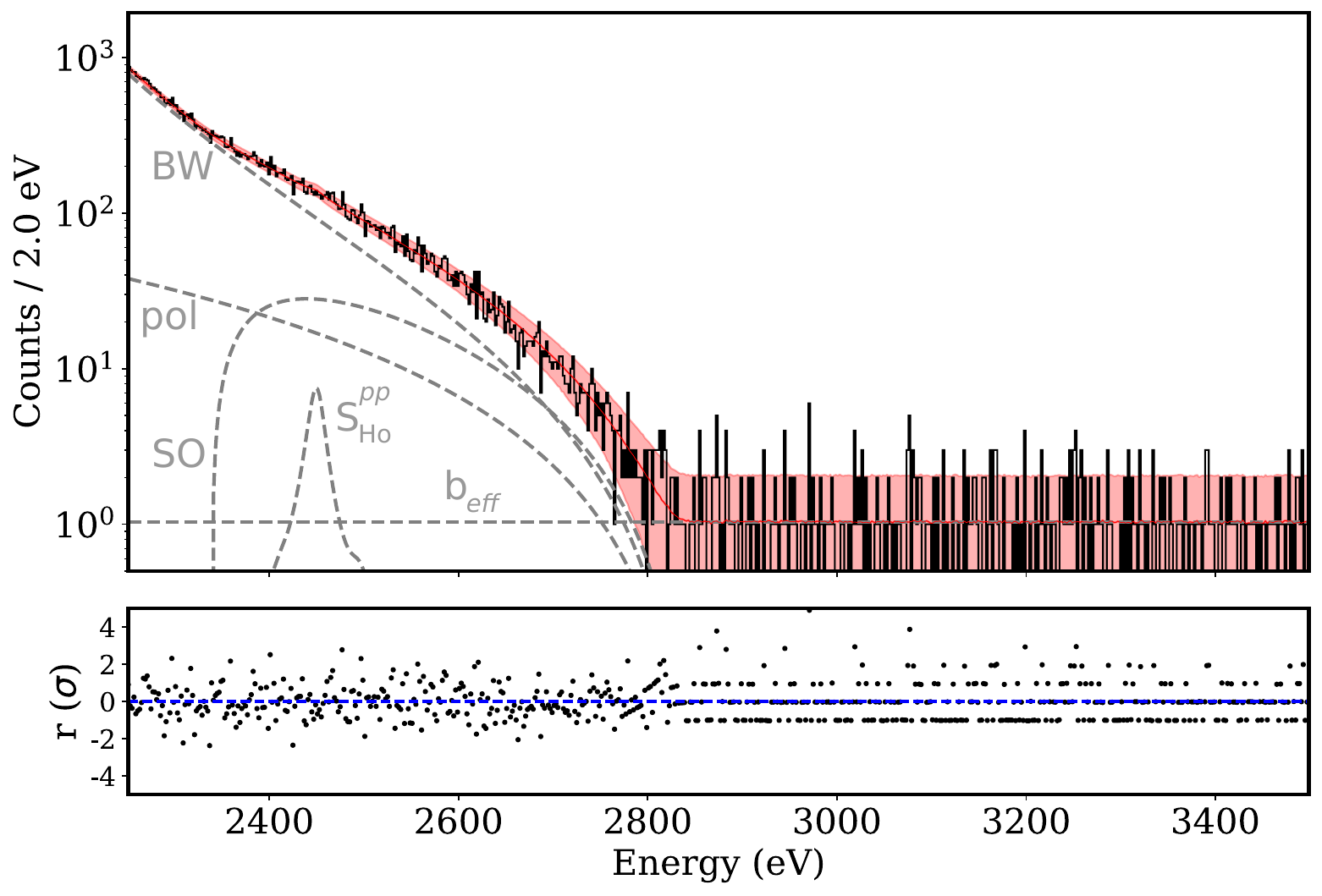}
\caption{\label{fig:mnuFit}
Top: Results of the Bayesian analysis of the calorimetric \Ho\ spectrum in the ROI with dashed lines showing the various components in Eq.\,(\ref{eq:SpecSum}) and Eq.\,(\ref{eq:HoEffect}). \refb{Each individual spectral component of Eq.\,(\ref{eq:HoEffect}) is multiplied by $\mathcal{F}_\n{PS}$ and convolved with $\mathcal{R}_{\rm eff}$.} The red line and the reddish band represent the mean and standard deviation of the distribution of the generated data, following the posteriors. The bottom part shows the residuals $r$ between the experimental data and the mean of the generated data, normalized by the standard deviation of the latter.}
\end{figure}

With the acquired statistics reported in this work, the ROI is chosen between 2250\,eV and 3500\,eV, where we find that the \Ho\ true spectrum $\mc{S}_\n{Ho}$ in Eq.\,(\ref{eq:SpecSum}) can be modelled as a sum of three terms (see also dashed lines in Fig.\,\ref{fig:mnuFit}):
\begin{equation}
\label{eq:HoEffect}
\mc{S}_\n{Ho} \approx \mc{S}^\prime_\n{Ho} = k_0 ( k_\n{BW} \mc{S}_\n{BW} + k_\n{SO} \mc{S}_\n{SO} + \mc{S}_\n{pol} ) \times \mathcal{F}_\n{PS},
\end{equation}
where $k_0$, $k_\n{BW}$ and $k_\n{SO}$ take care of the overall unit normalization.
$\mc{S}_\n{BW}$ describes the right tail of the M1 line
\begin{equation}
\label{eq:bw}
\mc{S}_\n{BW}(E_c| \gamma, E_{\mathrm{M1}}) =\frac{1}{2\pi} \frac{\gamma}{(E_c-E_{\mathrm{M1}})^2+\gamma^2/4},
\end{equation}
where $E_{\mathrm{M1}}$ and $\gamma$ are the line position and FWHM, respectively.
$\mc{S}_\n{SO}$ describes the energy spectrum of a shake-off de-excitation \cite{derujula_calorimetric_2016,brass_initio_2020}, parametrized as
\begin{multline}
\mc{S}_\n{SO}(E_c| E_{\n{so}}, \tau_1, \tau_2) = \\
 = \frac{1}{\tau_2-\tau_1}\Bigl( e^{-(E_c-E_\n{so})/\tau_2} -e^{-(E_c-E_\n{so})/\tau_1} \Bigr),
\end{multline}
where $E_\n{so}$, $\tau_1$ and $\tau_2$ are the shake-off transition energy, and the double exponential constants respectively.
$\mc{S}_\n{pol}(E_c)$ is a low degree polynomial, accounting for the tails of other peaks and shake-offs of the \Ho\ spectrum which are out of the ROI. Indeed, we find that just a constant term $\theta_0$ is enough, $\mc{S}_\n{pol}(E_c| \vec{\theta}) \simeq \theta_0$.
Finally, $\mathcal{F}_\n{PS}$ is the decay phase space factor, which is the only term that explicitly contains $m_{\beta}$
\begin{equation}
\label{eq:phasespace}
\mathcal{F}_\n{PS}(E_c| m_{\beta}, E_0) = (E_0-E_c)\sqrt{(E_0-E_c)^2-m_{\beta}^2}
\end{equation}

To extract the $m_{\beta}$ for the electron neutrino, we perform a Bayesian parameter estimation in the ROI using a Poisson likelihood (see Appendix D for details on parameter priors and fitting procedure) with the spectrum described by Eq.\,(\ref{eq:SpecSum}) and Eq.\,(\ref{eq:HoEffect}). The posterior is explored through a Hamiltonian Markov chain Monte Carlo using STAN \cite{standevelopmentteam_stan_2024}.
There are 13 free parameters in the fit. Among these, only 10 can be constrained by the data in the chosen ROI, namely $N$, $k_\n{BW}$, $k_\n{SO}$, $E_0$, $m_{\beta}$, $E_{\mathit{so}}$, $\tau_1$, $\tau_2$, $b_{\mathit{eff}}$, $\theta_0$ (as shown in Fig.\,\ref{fig:PriorPost} in Appendix\,D).

\begin{table}[tb!]
\caption{\label{tab:Pearson} Pearson correlation coefficients between key parameters in the Bayesian fit.}
\begin{flushleft}
\begin{ruledtabular}
\begin{tabular}{@{}cccccccc@{}}
 & \multicolumn{1}{l}{$E_0$} & \multicolumn{1}{l}{$m_{\beta}$} & \multicolumn{1}{l}{$\theta_0$} & \multicolumn{1}{l}{$N$} & \multicolumn{1}{l}{$k_\n{BW}$}            & \multicolumn{1}{l}{$k_\n{SO}$} & $\gamma$                  \\
\colrule
$m_{\beta}$ & 0.40                            & -                               & -0.06                         & 0.00                             & -0.01                                & -0.07                              & \multicolumn{1}{c}{0.00}     \\
$E_0$       & -                               & 0.40                            & -0.16                          & 0.04                          &  0.00                                 & -0.20                              & \multicolumn{1}{c}{0.00}
\end{tabular}
\end{ruledtabular}
\end{flushleft}
\begin{flushleft}
\begin{ruledtabular}
\begin{tabular}{@{}cccccccc@{}}
            & \multicolumn{1}{l}{$E_{\mathit{so}}$} & \multicolumn{1}{l}{$\tau_2$} & \multicolumn{1}{l}{$\tau_1$}   & \multicolumn{1}{l}{$b_{\mathit{eff}}$} & \multicolumn{1}{l}{$\Delta E_{\mathit{eff}}$} & \multicolumn{1}{l}{$N_{pp}$} &                           \\
\colrule
$m_{\beta}$ & 0.07                               &   0.00                         &  -0.13                           & 0.00                             & 0.00                                 & 0.00                                  &                           \\
$E_0$       & 0.27                               & -0.01                           & -0.50                           & -0.23                          & 0.00                                    & 0.00                                  &                           \\
\end{tabular}
\end{ruledtabular}
\end{flushleft}
\end{table}
It is worth emphasizing that, after parameter estimation, the posterior of the parameter of interest, $m_{\beta}$, is not directly correlated with any of the parameters describing the \Ho\ spectrum, as shown in Table \ref{tab:Pearson}, with the exception of $E_0$ (see also Fig.\,\ref{fig:corr_mnu} \refa{and Fig.\,\ref{fig:Pearson} in Appendix\,C}). This is expected: the phase space factor $\mathcal{F}_\n{PS}$ is the only term which contains $m_{\beta}$ and the \Ho\ spectrum is smooth at the end-point.

%

Finally, Fig.\,\ref{fig:corr_mnu} shows the results of this fit procedure on the recorded data, which results in an upper limit for the electron neutrino mass of $m_{\beta}<27$\,eV$/c^2$ at 90\% credibility. The endpoint is measured to be $E_0 = 2848^{+7}_{-6}$\,eV, compatible with the value reported in \cite{schweiger_penning-trap_2024b}.

\begin{figure}[tb!]
\includegraphics[width=0.5\textwidth]{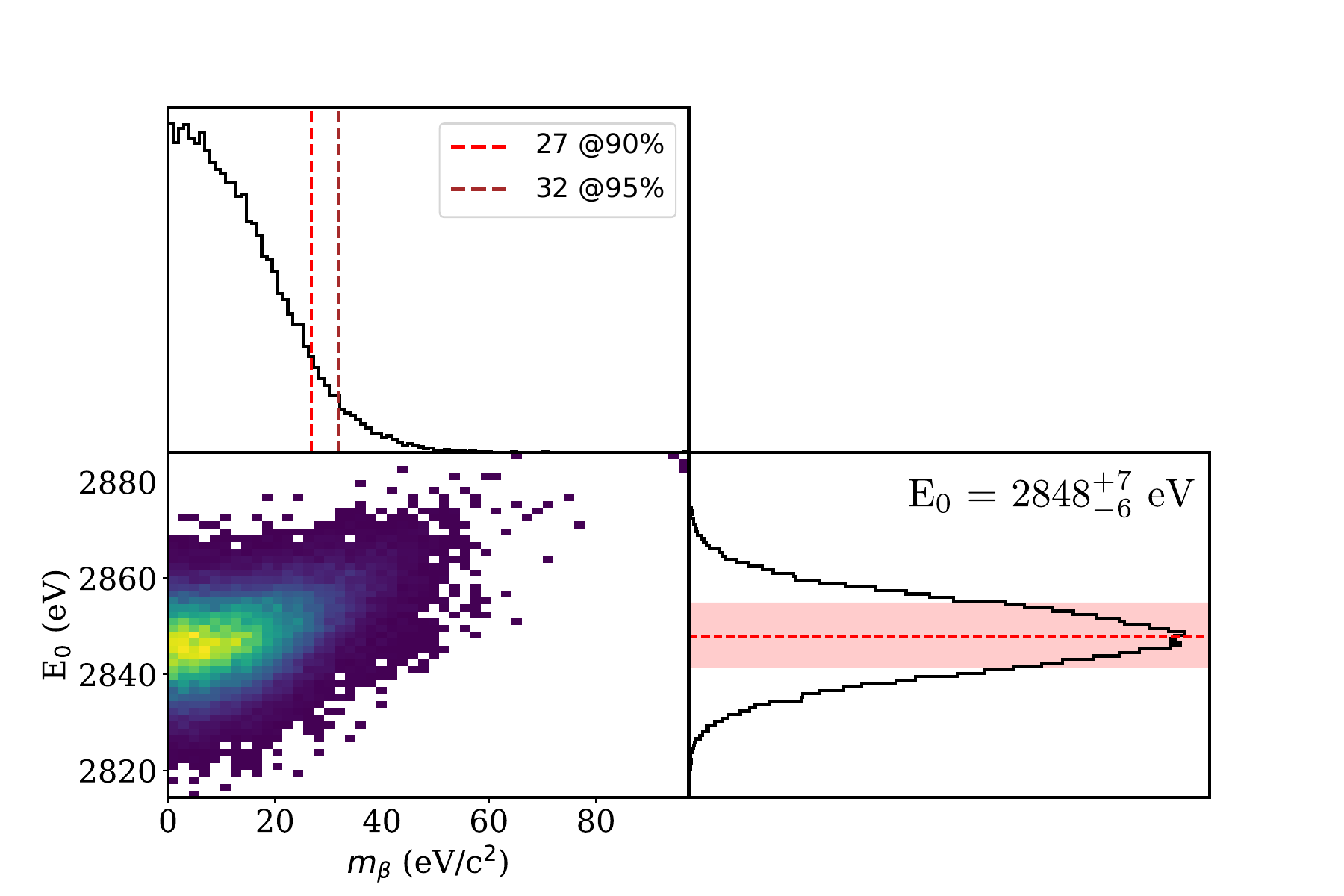}
\caption{\label{fig:corr_mnu}
Detail of the posteriors for $m_\beta$ and $E_0$ with their correlation as a result of the Bayesian analysis of the \Ho\ calorimetric spectrum.}
\end{figure}

The findings outlined here validate the approach first proposed more than 40 years ago in \cite{derujula_calorimetric_1982} and subsequently developed and implemented in recent years by the HOLMES and ECHo collaborations.
The new bound on the electron neutrino mass presented in this Letter is the strongest ever achieved studying the EC decay of \Ho, positioning the experiment as one of the most promising candidates for next-generation neutrino mass measurements.
\refa{
Although extending the neutrino mass sensitivity of this approach to the 0.1\,eV level requires increasing the overall statistics by a factor of about $10^9$, the favorable scalability of the critical experimental parameters of the prototype presented here -- namely, the number of detectors, \Ho\ activity, and measuring  duration -- renders this goal realistic for an experiment that  may be built up over time and distributed over multiple cryostats and institutions (see Appendix F for more details).
}
\refc{
However, it must also be recognized that as statistical sensitivity increases, systematic uncertainties -- especially those stemming from quantum and material origins -- will become increasingly significant and must be carefully addressed. In particular, uncertainties related to the chemical and crystalline environment of the holmium atoms will need to be thoroughly investigated and, if necessary, mitigated -- potentially through optimized detector fabrication processes.
}

Given that all key components have now been validated, further progress can be accelerated by leveraging modern microfabrication techniques to produce large-scale arrays with many thousands of detectors and by utilizing microwave multiplexing for the efficient readout of their signals,
thereby exploiting the scalability needed for next-generation experiments. The high-statistics spectrum recorded here also enables a realistic sensitivity study that will define the final configuration for a next-generation neutrino mass experiment, aiming for sub-0.1\,eV-scale sensitivity and opening the exploration of a range of neutrino masses that is presently inaccessible.

\begin{acknowledgments}
The HOLMES experiment has been supported by Istituto Nazionale di Fisica Nucleare (INFN) and by European Research Council under the European Union’s Seventh Framework Programme (FP7/2007–2013)/ERC Grant Agreement no. 340321.
\end{acknowledgments}

\appendix
\section{Appendix A: \Ho\ sample preparation and embedding}
\label{app:implant}
The \Ho\ isotope used in this work was produced by irradiating an \Er\ enriched Er$_2$O$_3$ sample with thermal neutrons in the high-flux nuclear reactor at the Institut Laue-Langevin (ILL, Grenoble, France). The produced \Ho\ was extracted from the irradiated sample using radiochemical methods \cite{heinitz_production_2018b}. However, the chemically purified sample still contains a fraction of the beta-decaying isomer \Hom\ (about $2\times10^{-3}$\,Bq(\Hom)/Bq(\Ho)) produced in the reactor and which must be removed to avoid excess background counts in the ROI. Isotope selection and ion implantation were performed by means of a dedicated system composed of a hot-running cold plasma sputter ion source coupled to a stirring magnet, a dipole magnet, and an adjustable slit.
\Ho\ ions exiting the ion source are accelerated to 30\,kV, selected by the dipole and further filtered by the slit, finally impinging on the microcalorimeter array with a current of about 5\,nA and a beam size of few millimeters FWHM \cite{degerone_development_2023a}.
The estimated separation of the \Ho\ and \Hom\ beams when they hit the array is about 6$\sigma$.
During ion implantation, the thick photoresist mask used to pattern the $180\times180$\,\mum$^2$ bottom gold layer was left in place to protect the rest of the array and was removed only after the deposition of the second gold layer.
In order to obtain an approximately uniform \Ho\ activity across the array, four implantation runs of about 3\,h each were performed with the array shifted between runs by several millimeters with respect to the beam center.
The \Ho\ ion current was monitored throughout implantation by measuring  the current flowing to ground through the gold layer covering the array.

\section{Appendix B: Detector array multiplexed readout}
\label{app:mux}
The signals of the 64 microcalorimeters are split and routed to two 32-channel rf-SQUID multiplexing chips, positioned along the two long sides of the microcalorimeter array (Fig.\,\ref{fig:holder}). These chips combine the frequency-converted signals into two 512\,MHz-wide bands, starting at 4\,GHz and 5\,GHz, respectively.  The multiplexed signals are transmitted through a single coaxial cable and amplified by a low-noise HEMT amplifier at 4\,K.  Signals from each multiplexing chip are recovered at room temperature using a Software-Defined Radio, implemented via two Reconfigurable Open Architecture Computing Hardware (ROACH2) boards \cite{mchugh_readout_2012}, in a heterodyne scheme. These boards, equipped with ADC/DAC modules, are combined with two Intermediate Frequency boards for up- and down-conversion. Software-triggered pulses are stored in a RAID system for further offline processing \cite{borghesi_first_2022}.

\section{Appendix C: Data analysis}
\label{app:ana}
\begin{figure}[tb!]
\includegraphics[width=0.45\textwidth]{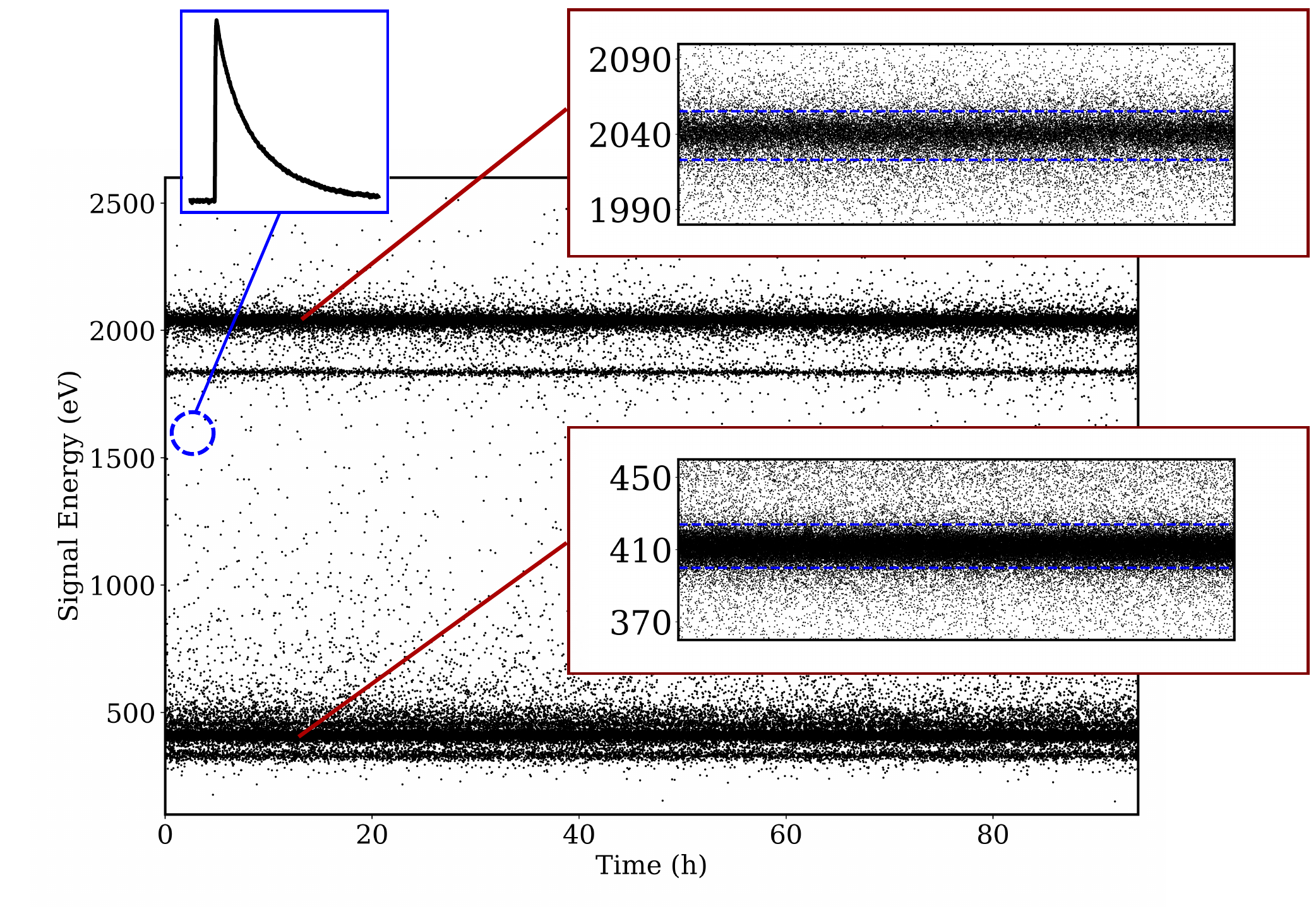}
\caption{\label{fig:stab}
Stability of the corrected energy gain over multiple days as shown by the events in the M and N peaks. The corrected gain drift remains well within the detector’s energy resolution minimizing systematic uncertainties in the energy scale.  The dashed lines in the insets on the right delimit a $\pm\sigma$ region around the mean of the highlighted peak. The inset on the left shows an impulse from a \Ho\ decay event.}
\end{figure}

Each of the about 1000 recorded raw spectra has been calibrated with a sequence of steps which includes \cite{borghesi_first_2022} 1) the rejection of spurious signals and too unstable time intervals, 2) the amplitude estimation by applying the optimal filter, 3) the gain time drift correction by monitoring the position of the M1, M2 and N1 peaks in the spectra, and 4) the energy calibration using the known positions of the same peaks.
The well designed cryogenic environment combined with the off-line analysis ensures a duty cycle of 82\% with a percentage of discarded events below 1\% and, as shown in Fig.\,\ref{fig:stab}, a corrected gain stability well within the detector energy resolution over a few days. The events discarded using mild linear cuts on pulse shape parameters \cite{borghesi_first_2022}, optimized for the ROI, are primarily signals distorted by pile-up and background radiation interacting with components of the microcalorimeters other than the detector absorber.

\section{Appendix D: Bayesian parameter estimation}
\label{app:post}
\begin{figure}[htb!]
\includegraphics[width=0.49\textwidth]{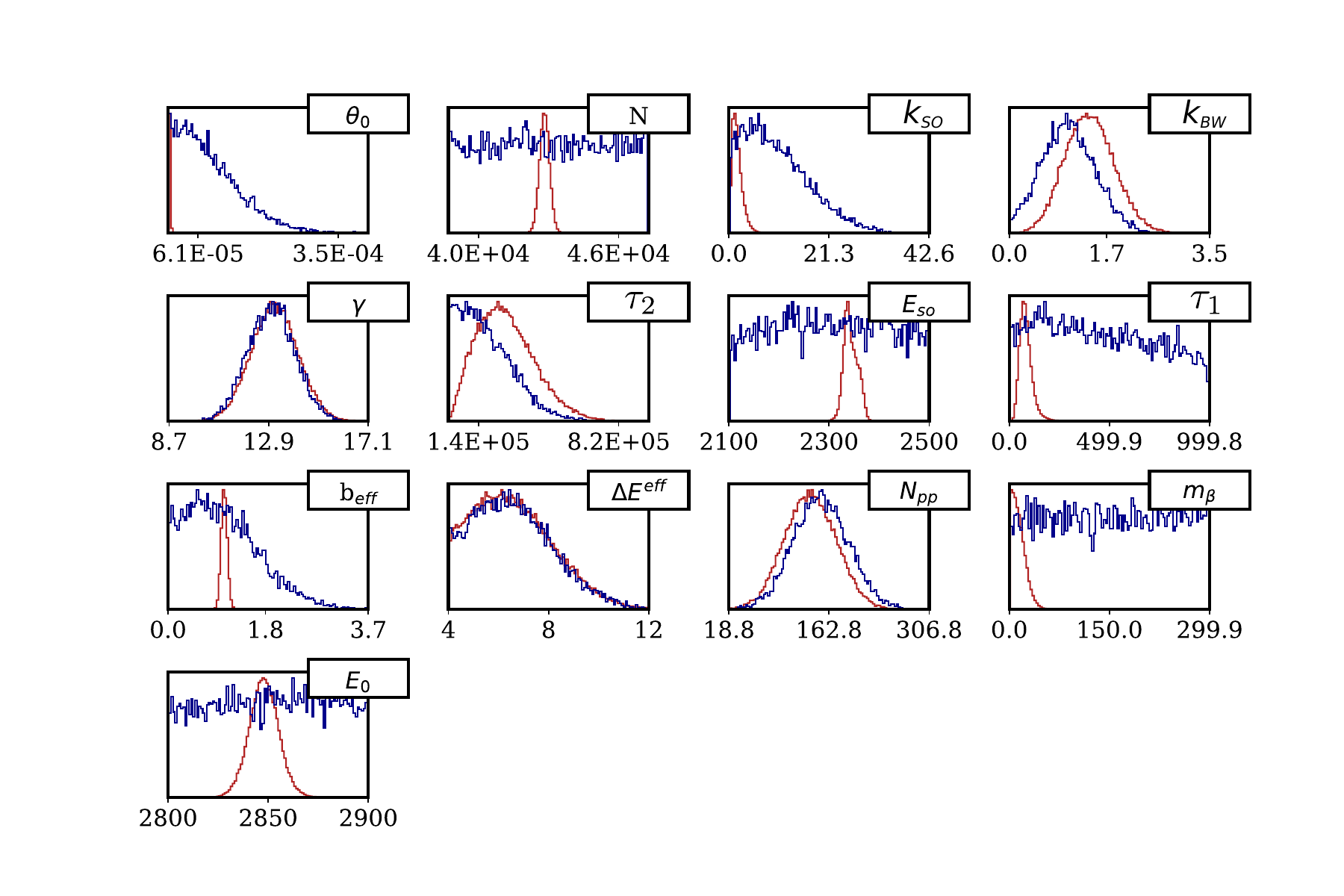}
\caption{\label{fig:PriorPost}
Prior (blue) and posterior (red) distributions for the key fit parameters used in the Bayesian analysis. The fit parameters include the endpoint
energy $E_0$, the neutrino mass $m_\beta$, and the other spectral shape parameters described in the text. }
\end{figure}

Bayesian fitting of the ROI using the model described by equations (\ref{eq:SpecSumTrue}) to (\ref{eq:phasespace}) involves estimating 13 parameters.
For the analysis, we normalize the spectra in the ROI so that, instead of $N_{tot}$ and $f_{pp}^{\mathit{eff}}$, we introduce $N$
and $N_{pp}$ which represent the number of decays and of pile-up events in the ROI, respectively.
Data in the chosen ROI can constrain only 10 of the 13 parameters, namely $N$, $k_\n{BW}$, $k_\n{SO}$, $E_0$, $m_{\beta}$, $E_{\mathit{so}}$, $\tau_1$, $\tau_2$, $b_{\mathit{eff}}$, $\theta_0$, as shown in Fig.\,\ref{fig:PriorPost}. 
For these parameters, we use uninformative prior distributions with large standard deviations, while
for the remaining 3 parameters, $N_{pp}$, $\Delta E_{\mathit{eff}}$, and $\gamma$, we use weakly informative priors. \refb{All priors are taken as normal distributions.}
The priors for the effective energy resolution $\Delta E_{\mathit{eff}}$ in Eq.\,(\ref{eq:SpecSumTrue}) and for $N_{pp}$ are set to allow variations within reasonable ranges around the values expected from the measured $\langle \Delta E_\n{FWHM}\rangle$ and the estimated $f_{pp}$, respectively.
The peak marked $\mc{S}_\n{Ho}^{pp}$ in Fig.\,\ref{fig:mnuFit} comes from the self-convolution of the N and M peaks in the pile-up spectrum and its amplitude is too low to be constrained by the data.
While the position of the M1 Breit-Wigner peak $E_{\mathrm{M1}}$ in Eq.\,(\ref{eq:bw}) is fixed for simplicity to the value, its FWHM ($\gamma$) is set to allow variation within a reasonable range. For both, we use the values obtained from our data as described above.
Although the $Q$-value of \Ho\ has been measured with high precision \cite{schweiger_penning-trap_2024b}, it is considered good practice to treat the endpoint $E_0$ of the spectrum as a free parameter. Errors in the energy determination of the main peaks used for calibrating the summed spectrum could shift the fitted endpoint energy. This shift is accurately accounted for only if the endpoint is allowed to vary.

\begin{figure}[htb!]
\includegraphics[width=0.4\textwidth]{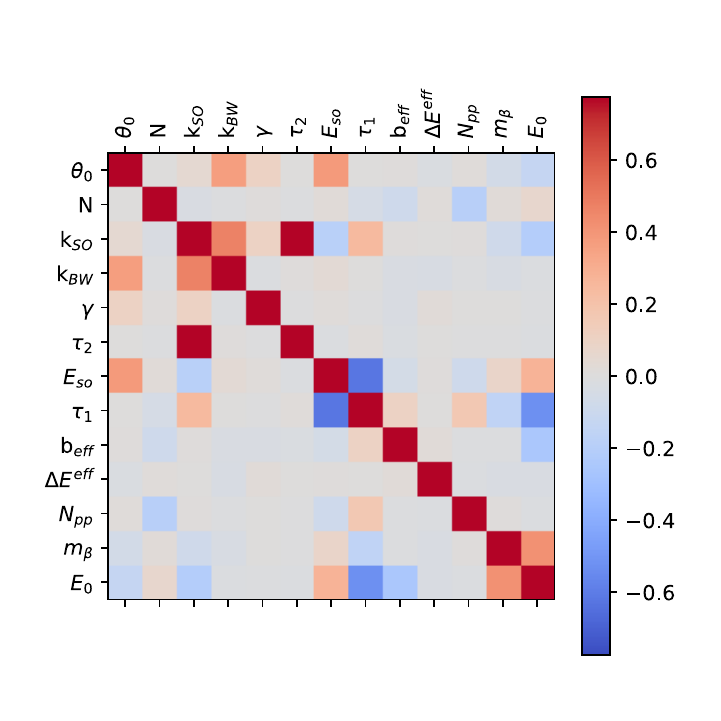}
\caption{\label{fig:Pearson}
Correlation matrix for the fitted parameters.}
\end{figure}
\refb{Figure\,\ref{fig:Pearson} shows the full correlation matrix (Pearson coefficients) for all fitted parameters, from which the values reported in Table\,\ref{tab:Pearson} are derived.}

To verify the fidelity of the effective model of Eq.\,(\ref{eq:HoEffect}) in describing the \Ho\ spectrum within the chosen ROI, we use Monte Carlo simulations.
A set of $n$ toy experiments is simulated. For each experiment the recorded data are resampled with statistical fluctuations, and each time a fit with the effective model (Eqs.\,\ref{eq:SpecSumTrue} and \ref{eq:HoEffect}) is performed to evaluate the upper limit for the 90\% credible interval of $m_\beta$.
The resulting distribution has a mean of 40\,eV$/c^2$ and a standard deviation of 10\,eV$/c^2$, which is compatible with the result obtained from the real data.

\section{\refa{Appendix E: Monte Carlo investigation of the relevant systematic effects.}}
\label{app:sys}
As mentioned in the article, with the level of statistics collected in this measurement, there are two possible sources of systematic effects that should be investigated. These are related to the summing of many data sets, each characterized by an uncertainty in the energy scale within the ROI and by a different Gaussian response.
The extrapolation of the calibration curve to the ROI, given the combined effect of the detector's non-linear response and the positions of the calibration peaks, could, in principle, cause a shift in the quadratic energy estimator $\hat{E}_i$ of each $i$-th dataset. 
This effect can be modeled as $\hat{E}_i = E_t + \delta E_i(E_t)$, where $E_t$ is the true event energy, 
and $\delta E_i(E_t)$ is the unknown energy shift, which is assumed to be linear in true energy, equal to zero at the last calibration peak (M1) and equal to $\delta_i$ at $Q$, i.e.,
\begin{equation}
    \delta E_i(E_t) =  \delta_i \frac{E_t - E_\mathrm{M1}}{Q-E_{\mathrm{M1}}} \nonumber
\end{equation}
Based on previous measurements, we expect $\delta_i/Q$ to be around 0.2\%. In other words, for each of the approximately 1000 datasets, the energy spectrum in the ROI could be shifted upward or downward by less than 10\,eV.

To investigate both this effect and the use of the average Gaussian response $\mc{R}_\im{eff}(E_c) \simeq \mathcal{G}(E_c |0, \Delta E_{\mathit{eff}})$, we generated $\mathcal{O}(100)$ Monte Carlo spectra, each convoluted with a single Gaussian resolution of 6\,eV and no energy shift. We then performed parameter estimation on each of them. The resulting distribution of the 90\% upper limits on $m_\beta$ serves as our reference.

Next, we simulated $\mathcal{O}(100)$ Monte Carlo spectra, each composed of the sum of 1000 different spectra with varying Gaussian energy resolutions and energy scale shifts $\delta E_i(E_t)$. The energy resolutions were distributed as in the inset of Fig.\,\ref{fig:spettro}, and $\delta_i=\delta E_i(Q)$ were conservatively normally distributed around 0\,eV with a standard deviation of 10\,eV. We then performed the same parameter estimation as before, using the approximated formula in Eq.\,(\ref{eq:SpecSum}), and compared the resulting distribution of 90\% upper limits with the reference one.

Since no significant difference is observed, with the latter (reference) distribution showing a mean of 42\,eV (44\,eV) and a standard deviation of 10\,eV (10\,eV), we can safely conclude that these effects are negligible given the current acquired statistics.

\section{\refa{Appendix F: Scaled sensitivity of future holmium-based experiments}}
\label{app:scaling}

A forthcoming publication is in preparation to investigate, through detailed Monte Carlo simulations, the optimal experimental configuration required to achieve sub-0.1\,eV-scale statistical sensitivities. Nevertheless, it is worth briefly elaborating on the approximate scaling factor mentioned in the conclusions of this work. The $10^9$ factor is an order-of-magnitude estimate for achieving the target 0.1\,eV sensitivity, based on the expected scaling of the statistical sensitivity with $N^{-1/4}$ \cite{nucciotti_expectations_2010a,nucciotti_statistical_2014b}, where $N$ is the number of decays, analogous to other neutrino mass endpoint measurements. This scaling is approximate, as it does not account for the beneficial effects of improving energy resolution or reducing radioactive background. At the same time, it also neglects the adverse impact of increased pile-up levels associated with higher detector activity. With these caveats in mind, we can provide one possible breakdown of the scaling factor. The required $10^9$-fold increase in statistics could be achieved by increasing the detector count by about $10^4$ (i.e., to approximately $10^6$), the \Ho\ activity per detector by about $10^3$ (i.e., to the order of 100\,Bq), and the measuring time by about $10^2$ (i.e., to approximately 10 years). 

Such an experiment could be deployed in a few commercial dilution refrigerators at different experimental sites, including underground laboratories if needed to minimize radioactive background. 
While this approach would add reliability and flexibility, it is worth noting that a single installation, such as the large high-power dilution refrigerator developed and used for the CUORE neutrinoless double beta decay experiment \cite{adams_cuore_2022a, adams_search_2022b}, would be sufficient to host the final experiment. The CUORE cryogenic system has provided a 1-cubic-meter experimental volume at temperatures below 10\,mK, with heavy shielding against environmental background, since 2017. Similar refrigerators are being prepared for other rare event searches and for operating large Quantum Processing Units with superconducting qubits \cite{agrawal_projected_2025,mohseni_how_2024}.
\refc{It is also important to note that, for future neutrino mass experiments, the necessity of operating in underground sites should be carefully evaluated based on the identified background sources and only after all alternative mitigation strategies have been considered. For example, active background rejection could be achieved by implementing anticoincidence detector channels fabricated directly on the silicon substrate of each detector array chip \cite{ferraribarusso_status_2025}.}

To prevent the cost of a future experiment from becoming prohibitive, it will be necessary to leverage emerging multiplexing schemes, advanced signal processing electronics for telecommunications and quantum computing, as well as more efficient ion beam sources for implantation.

In conclusion, while these technical improvements are challenging, they are shared with many other fields and have already been extensively demonstrated in various real-world applications.

While the developed technology is claimed to have the potential for scaling up to the required size, it should be emphasized that this process will be gradual and is expected to take at least a decade. During this period, despite the advantages of the calorimetric approach, new sources of systematic uncertainty must be addressed and thoroughly investigated. Naturally, it cannot be entirely ruled out that unforeseen effects, emerging as the statistical sensitivity improves with the larger scale of the experiments, could present significant challenges.
\refc{
Two main classes of challenges can already be anticipated. First, the substantial increase in experimental scale will make accurate raw data reduction increasingly demanding, with potential systematic errors arising, for example, from uncertainties in determining the energy resolution at the endpoint for each partial data set, and from combining a large number of data sets with some spread in the Gaussian response. Second, the enhanced statistical sensitivity will require a precise investigation of all potential quantum effects due to the chemical and lattice environment of \Ho.
}

\bibliographystyle{apsrev4-2}
\bibliography{new_limit}

\end{document}